\documentclass[aps,twocolumn,amsmath,amssymb,superscriptaddress]{revtex4}

\usepackage[latin1]{inputenc}
\usepackage{graphics}
\usepackage{dcolumn}
\usepackage{graphicx}
\usepackage{dcolumn}
\usepackage{bm}
\usepackage{graphics}
\usepackage{amsmath,amssymb}
\providecommand{\eqref}[1]{(\ref{#1})}
\providecommand{\erf}{\operatorname{erf}}
\providecommand{\qperpv}{\ensuremath{\vec{q} \! \perp \! \vec{v}}}
\providecommand{\qparav}{\ensuremath{\vec{q} \! \parallel \! \vec{v}}}
\providecommand{\microns}{\ensuremath{\mu\text{m}}}
\providecommand{\vq}{\ensuremath{\vec{q}}}
\providecommand{\vv}{\ensuremath{\vec{v}}}
\begin{document}
\title{Dynamics in shear flow studied by 
X-ray Photon Correlation Spectroscopy}
\author{Sebastian Busch} 
\affiliation{Physik Department, TU M\"unchen, Garching bei M\"unchen, Germany}
\author{Torben Haugaard Jensen}
\affiliation{Niels Bohr Institute, University of Copenhagen, Denmark}
\author{Yuriy Chushkin}
\author{Andrei Fluerasu}
\affiliation{European Synchrotron Radiation Facility, Grenoble, France}

\date{\today}

\begin{abstract}
X-ray photon correlation spectroscopy was used to measure the diffusive 
dynamics of colloidal particles in a shear flow. The results 
presented here show how the intensity autocorrelation functions measure 
both the diffusive dynamics of the particles and their
flow-induced, convective motion. However, in the limit of low flow/shear 
rates, it is possible to obtain the diffusive component of the dynamics, 
which makes the method suitable for the study of the dynamical properties
of a large class of complex soft-matter and biological fluids. 
An important benefit of this experimental strategy over more traditional 
X-ray methods is the minimization of X-ray induced beam damage. While the
method can be applied also for photon correlation spectroscopy in the
visible domain, our analysis shows that the experimental conditions under
which it is possible to measure the diffusive dynamics are easier to achieve
at higher q values (with X-rays).
\end{abstract}
\maketitle
\section{Introduction}
\label{intro}

X-ray photon correlation spectroscopy (XPCS) offers an unique way to 
perform direct measurements on the meso\-scale dynamics in a large class of 
complex soft-matter systems (see for example Refs. 
\cite{gel_PRE_07,Falus_PRL97_2006,Bandyopadhyay_PRL04,Lurio_PRL00,Lumma_PRE00,Banchio_PRL06}).
Based on the same principle as Dynamic Light Scattering (DLS) 
\cite{Berne_Pecora}, XPCS is however not limited to non-turbid samples, multiple scattering can usually be neglected, and can access larger values 
of the momentum transfer \vq, probing smaller spacial distances. 
Two major difficulties associated with XPCS are the small scattering 
strength of most samples and their degradation induced by the X-rays. 
Using a high-brilliance $3^\text{rd}$ generation synchrotron source helps 
addressing the former issue but increases at the same time the importance 
of the latter. In the case of fluid samples, performing XPCS experiments 
under continuous flow can limit the effects of beam damage. In addition, fluid 
mixers, in which the temporal dependence of a process taking place in a 
flow device is mapped into a spatial dependence along the flow channel,
can be used to perform time-resolved studies \cite{Pollack_PRL01,microfluidics}.

In the experiments reported here, XPCS was used to study the Brownian 
dynamics of a colloidal suspension of (weakly to non-interacting) hard 
spheres under shear flow. 
In general, the correlation functions measured by XPCS are affected by both, 
the flow-induced dynamics as well as the diffusive dynamics of the particles. 
While measuring the flow velocity profile can be a legitimate goal in many 
applications (see for example Ref.~\cite{Narayan_AO97}), the aim of the
present experiment was to describe the diffusive component of the dynamics. 
The results presented here show that for a laminar flow, and in the limits
of low shear rates, it is possible to decouple the components due to
diffusive and convective motion of the particles. 

\section{Description of the experiment}
\label{sec:Experiments}

\subsection{X-ray photon correlation spectroscopy}

\begin{figure}
  \resizebox{0.7\columnwidth}{!}{%
  \includegraphics{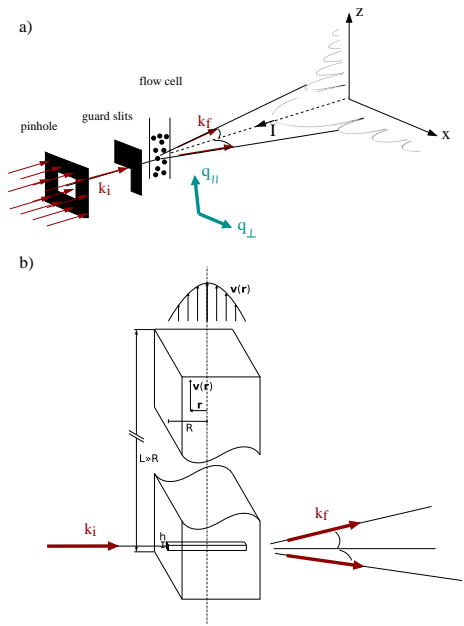}} 
  \caption{(a) XPCS experimental setup for the flow geometry considered here. 
Longitudinal flow scans were performed by moving the detector in the 
vertical $z$ direction, while transverse flow scans were performed by 
keeping the detector in the horizontal scattering plane and moving it along 
the $x$ direction. 
(b) Schematics of the flow channel with quadratic section 
(edge length $2R$). The velocity 
profile has a parabolic shape; $h$ is the transverse size of the X-ray beam.}
\label{fig:flowcell}
\end{figure}

The experiment was performed at the Tro\"ika beamline (ID10A) of the 
European Synchrotron Radiation Facility (ESRF) in Grenoble, France. 
A single bounce Si(111) crystal monochromator was used to select 8\,keV 
X-rays, corresponding to a wavelength of $\lambda = 1.55\,\text{\AA}$. 
The relative bandwidth was $\Delta \lambda /\lambda \approx  10^{-4}$. 
A Si mirror downstream of the monochromator was suppressing higher order 
frequencies.  
The source size (FWHM) was approximately $30\times900\,\microns^2$ 
(v$\times$h) and the source-sample distance was 46\,m. In the vertical 
direction, the X-ray beam was focused by a beryllium compound refractive 
lens in order to enhance the flux. A transversely partially 
coherent beam was defined by slit blades with highly polished cylindrical 
edges. The slit size was 10\,\microns\ in the vertical and horizontal 
directions. A set of guard slits, placed just upstream of the sample was 
used to minimize the parasitic X-ray scattering from the beam-defining slits
(see Fig.~\ref{fig:flowcell}a). 
Under these conditions, the incident X-ray flux was 
$\sim5\cdot10^9$~photons/s. 
The scattering from the colloidal suspension was recorded with a Cyberstar 
scintillator point detector placed at a distance of 2.30\,m from the sample. 
The detection area was typically limited to $100\times100\,\microns^2$ by
precision slits placed in front of the detector, 
corresponding to a few speckles. 

As it will be seen, the dynamics measured in a flowing sample 
is not isotropic. In the experiments described here,  
measurements were performed at 
momentum transfers covering a range of $q\cdot a$ from 1.5 to 5.7 
($a$ is the radius of the particles) in ``transverse flow'' and 
``longitudinal flow'' geometries (Fig.~\ref{fig:flowcell}a).

Normalized intensity fluctuation autocorrelation functions,
\begin{equation}
  g_2(\vq,t) = \frac{\left< I(\vq,t_0) \cdot I(\vq,t_0+t) \right>_{t_0}}{\left< I(\vq,t_0)\right>_{t_0}^2 } ,
\label{eq:g2}
\end{equation}
were obtained using a ``Flex'' hardware correlator manufactured by 
Correlator.com that was connected to the output of the X-ray detector.

\subsection{Flowcell}

The in-house made flowcell consisted of an aluminum sheet through which 
the flow channel (see sketch in Fig.~\ref{fig:flowcell}b) was cut by 
high-precision machining. Polymer-coated kapton foils were fusion-bonded to 
the face of the aluminum and held in place by a ``sandwich-type'' pressure 
applying system. A high-precision syringe pump obtained from Harvard Apparatus 
was used to push the fluid through the flow device. The width of the channel 
($2R = 1000\,\microns$) was much larger than the one of the beam 
($h = 10\,\microns$). Therefore, the scattering volume could be approximated 
by a one-dimensional object. This made it possible to apply the model of a 
Poiseuille flow -- a Newtonian fluid performing a shear flow in a 
cylindrically-shaped flow device.

The laminar character of the flow is determined by the 
\emph{Reynolds number} which measures 
the ratio of the inertial to viscous forces, and is defined as 
$Re = \frac{v_0 2R \rho}{\eta}$ \cite{microfluidics}. Here $2R$ is the 
characteristic length of the system (confer to Fig.~\ref{fig:flowcell}b), 
$v_0$ is the flow velocity of the fluid and $\eta$ the dynamic viscosity. 
As long as the Reynolds numbers are low ($Re \ll 10^3$), the flow is 
in a laminar regime, which was well the case in all the measurements performed 
here, as can be 
seen in table~\ref{tab:samples}. Assuming a no-slip boundary condition at 
the wall of the flow cell, the velocity profile can therefore be described 
by
\begin{equation}
  v(r) = 2 v_0 \left( 1 - \frac{r^2}{R^2} \right) .
\label{eq:vf}
\end{equation}
Here (see Fig.~\ref{fig:flowcell}b), $r$ is the distance from the center 
of the capillary and $v_0$ is the average flow velocity (defined using 
the average flow rate $Q = 4 R^2 v_0$). The mean shear rate is defined
as $\dot{\gamma} = \frac{2 v_0}{R}$.

\subsection{Sample}

Charge stabilized latex spheres with a nominal radius of $a$=110\,nm dispersed 
in water were obtained from Duke Scientific Co.\ at a concentration of 
10\% per weight. Glycerol with a purity of $\geq$99\% was obtained from 
Sigma-Aldrich Co. The spheres suspension was mixed with the glycerol and 
placed in a dessicator which was evacuated on top of a heatable magnetic 
stirrer. The temperature was set to approximately 75$^\circ$C which 
facilitated water evaporation and decreased the viscosity. These 
conditions were applied for about two days, leaving the samples not only 
essentially water-free but also de-gased, avoiding thus an important reason 
for bubble formation. The amount of suspension and glycerol was chosen such 
that the final volume fraction was $\Phi = 0.1$.

\begin{figure}
\resizebox{0.6\columnwidth}{!}{%
  \includegraphics{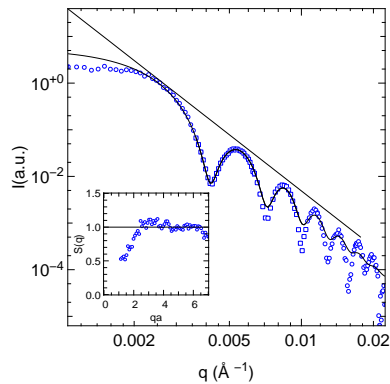}
}
\caption{Static SAXS scattering from the latex particles and fitted
form factor for the solid spherical particles (solid line). 
The straight line with slope -4, representing Porod's law 
is a guide for the eyes.
Inset: estimate of the structure factor $S(q)$ obtained by dividing 
the total scattered intensity to the fitted value for the form factor.
}
\label{fig:saxs}       
\end{figure}

The small angle X-ray scattering (SAXS) pattern of the sample is shown in 
Fig.~\ref{fig:saxs}. The time-averaged data was fitted, for values of 
$qa\geq 5.0$, to the form factor of spherical particles with the size 
distribution given by a Schultz distribution function, 
yielding a radius of $a = 1070\,\text{\AA}$ and a size polydispersity $\leq$ 7\%. 
Also shown is the static structure factor $S(q)$, exhibiting the weakness of 
interactions ($S(q)\approx 1$ for most of the $q$-range).

\section{Theory: XPCS in a laminar flow}
\label{sec:Theory}

The central quantity describing the dynamics of the colloidal suspension is the 
dynamic structure factor, or intermediate scattering function (ISF), which in 
the case of non-interacting (statistically independent) particles is given by 
\cite{Berne_Pecora}
\begin{equation}
  g_1(\vq, t) \propto \sum_{k=1}^N \left< E_k^*(0)E_k(t) \exp\left[ -i \vq \cdot \big( \vec{r'}_k(0) - \vec{r'}_k(t) \big) \right] \right>
  \label{eq:ISF}.
\end{equation}
Here {\bf q} is the scattering vector, $E_k(t)$ is the amplitude of X-rays 
scattered by particle $k$ at time t and $\vec{r'}_k(t)$ is the particle 
position in the presence of flow at time t. The summation is over all 
particles in the scattering volume for a particular realization of
the experiment and the ensemble average is performed over many different 
realizations. In order to differentiate between the diffusive and the 
flow-induced particle displacements, it is useful to write
\begin{multline}
\exp\left[ -i \vq \cdot \big( \vec{r'}_k(0) - \vec{r'}_k(t) \big) \right] = \\
	\exp\left[ -i \vq \cdot \big( \vec{r}_k(0) - \vec{r}_k(t) \big) \right] \cdot 
	\exp (i \vq \cdot \vv_0 t) \cdot \exp(i\vq \cdot \delta \vv t)
  \label{eq:diff_flow}.
\end{multline}
Here $\vec{r}_k(t)$ is the particle position due to only to the diffusive motion, $\vec v_0$
is the average flow velocity throughout the sample and $\delta \vec{v}$ is the flow 
velocity difference between positions $\vec{r}_k(t)$ and $\vec{r}_k(0)$.

By analyzing Eqs.~\ref{eq:ISF} and~\ref{eq:diff_flow}, it can be 
concluded that the ISF depends on three main factors:\\
i) the particle diffusion, with a characteristic time scale 
$\tau_D=(Dq^2)^{-1}$ 
($D$ is the diffusion coefficient, see below)\\
ii) the amplitude factor, which measures the intensity scattered by each particle,
$I_k=\left<E^*_k(0)\cdot E_k(0)\right>$, but also the transit time of the particles 
through the scattering volume, $\tau_T=h/v_0$ ($h$ is the transverse beam size),\\
iii) shear-induced effects which are described, as it will be shown, by a 
time scale $\tau_S \propto (q \dot \gamma R \cos\phi )^{-1}$, which is 
inversely proportional to the magnitude of the local velocity 
gradient $\dot\gamma$. Here $\phi$ is the angle between the 
scattering vector and the local velocity $\vec{v}$.

Note that the importance of the fourth factor, namely the average velocity-dependent 
$\exp \left[i \vq \cdot \vv_0 t \right]$, was not emphasized because it 
can only be detected in heterodyne correlation functions \cite{mark_heterodyne_07}.
A homodyne PCS experiment (like the one reported here) 
detecting the square modulus of the ISF, 
$\left|g_1({\bf q}, t)\right|^2$ can measure velocity gradients but not 
absolute velocities
\cite{JB_EPJAP}.

It can be demonstrated \cite{Fuller} that, if the three time scales described
above are sufficiently different from each other, the ISF can be factorized in
independent contributions from the three relaxation mechanisms. The physical 
argument is that if one of the contributions 
(e.g. diffusion) to the ISF varies on a much faster time scale than the 
others, then the particles explore the phase space associated with 
this particular relaxation mechanism while the other factors are 
(approximately) constant. In such a situation, the ISF
under shear flow can be written as
\begin{multline}
  g_1(\vq, t) \propto \exp(i \vq \vv_0 t) \left< E^*(0)E(t)\right> \times \\
\left< \sum_{k=1}^N 
\exp\left[ -i \vq \cdot \big( \vec{r}_k(0) - \vec{r}_k(t) \big) \right] \right> 
\times \\
\left< \sum_{k,l=1}^N \exp \left[i \vq \cdot \delta \vv(k,l) t \right] \right>
  \label{eq:ISF_fact}.
\end{multline}
In writing Eq.~\ref{eq:ISF_fact}, we have also assumed that all particles 
are identical scatterers and that, due to the fact that particles are 
indistinguishable and that the system is stationary, the shear induced 
effects can be calculated also by taking into account the velocity
differences between all pair of particles for a particular realization of the 
statistical ensemble.

The physical interpretation of the shear-induced effects is that
the frequencies of X-rays scattered by particles which are moving with different
(average) velocities are Doppler-shifted with respect to each other, and 
produce a beat frequency when interfering on the detector. For a pair 
of particles situated at two different locations and moving with 
a velocity difference of $\delta\vec{v}$, the self-beat frequency  
induced by the Doppler shifts is $\vq \cdot \delta \vec{v}$ 
\cite{Berne_Pecora,Narayan_AO97,Fuller}.
From here, it can be seen that measurements with \qparav\ will be influenced 
by these velocity-differences while measurements with \qperpv\ are not 
affected because the scalar products $\vq \cdot \vec{\delta v}$ are
all zero. 

In the case of practical interest for our study, namely the situation in which
the diffusion relaxation time is much shorter than the other relevant time scales,
it is natural to assume that the scattered intensity fluctuations are well
described by Gaussian statistics and that the intensity autocorrelation functions 
are related to the intermediate scattering functions by the Siegert relationship,
\begin{equation}
  g_2(\vq, t) = 1 + \beta \cdot \left| g_1(\vq, t) \right|^2 ,
  \label{eq:sig}
\end{equation}
where $\beta$ is the speckle contrast which was, in the described 
experimental setup, of the 
order of 5\,\%. In ref. \cite{Fuller} it is also mentioned that ``it can be shown''
that this relation is also valid for the case where the flow-induced deterministic motions 
dominate those due to diffusion, provided that the decay of the ISF is sufficiently short 
compared to the inverse local velocity gradient $\dot\gamma^{-1}$. 
This condition is almost always fulfilled with typical experimental 
parameters, and in particular is satisfied for 
our longitudinal flow scattering geometry (\qparav), which justifies the 
use of the Siegert relationship in all the experiments performed here. 

In conclusion, the homodyne correlation functions measured in our XPCS experiments
can be well described by, 
\begin{equation}
  |g_1(\vq,t)|^2 = |g_{1,D}(\vq,t)|^2 \cdot |g_{1,T}(t)|^2 \cdot |g_{1,S}(\vq,t)|^2 ,
  \label{eq:g1tot}
\end{equation}
where the first factor (subscript~$_D$) is due to thermal diffusion, the second one
(subscript~$_T$) to the transit time through the scattering volume, and the last one 
(subscript~$_S$) is due to shear. Each of these factors are discussed in the following.

\subsection{Thermal diffusion}

The intermediate scattering function for a non-flowing sample undergoing
Brownian motion is a simple exponential decay with the relaxation rate
determined by the diffusion coefficient $D$ and scattering wavevector 
{\bf $q$},
\begin{equation}
  |g_{1,D}(\vq, t)|^2 = \exp \left[ - 2 D q^2 t \right]
  \label{eq:g1diff}.
\end{equation}
In this case, the dynamics is isotropic and does not depend on the direction
of ${\bf q}$, but only on its absolute value $q$.
When a uniform shear rate $\dot{\gamma}$ is applied, the diffusion is in
general enhanced
by the shear and becomes anisotropic, 
changing the formula to \cite{Ackerson_Clark_JPhysique81}:
\begin{equation}
  |g_{1,D}(\vq, t)|^2 = \exp \left[ - 2 D q^2 t
   \left( 1 - \frac{q_\parallel q_\perp}{q^2} \dot{\gamma}t +   
    \frac{1}{3}\frac{q_\parallel^2}{q^2} (\dot{\gamma}t)^2 \right) \right] .
  \label{eq:g1D}
\end{equation}
Here, $q_\parallel$ and $q_\perp$ are the components of \vq\ parallel and respectively perpendicular to the direction of the flow. 
In a transverse scattering geometry (\qperpv), $q_\parallel$=0 and 
Eq.~\eqref{eq:g1D} reduces to Eq.~\eqref{eq:g1diff}. 
This is obviously not true when \qparav, in which case the diffusion 
is enhanced by the shear. 
However, at the shear rates used here 
(see table~\ref{tab:samples}), the products $t\dot{\gamma}$ are of the order 
of 10$^{-3}$ and Eq.~\eqref{eq:g1D} provides only a minor correction to
Eq.~\eqref{eq:g1diff}, which can still be considered to describe well 
the thermal diffusion of the colloids.

\begin{figure}
\resizebox{0.6\columnwidth}{!}{%
  \includegraphics{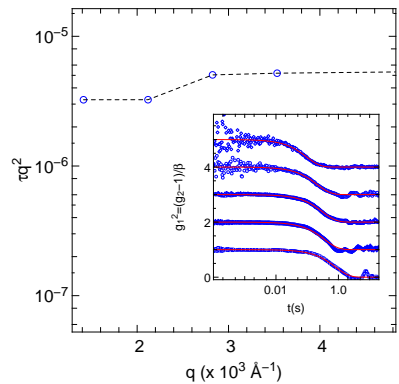}
}
\caption{Correlation functions taken at no flow, $Q = 0$ (sample in 
a capillary) at all values of the scattering wavevector measured, 
$q$= 1.8, 2.1, 2.8, 2.5, and 5.3 $\times$10$^{-3}$ \AA$^{-1}$. The inset 
shows the normalized correlation functions shifted vertically, for clarity, 
by 0, 1, 2, 3, and respectively 4, and fits with exponential 
decays - Eq.~\eqref{eq:g1diff}, solid lines. The main graph shows the 
correlation times $\tau=1/Dq^2$ (multiplied by $q^2$ in order to
eliminate the q-dependence) as a function of $q$.}
\label{fig:noflow}       
\end{figure}

\subsection{Transit-time effects}

When the colloidal suspension flows through the beam, new particles enter 
the scattering volume, replacing particles that 
leave on the other side. 
The ``refilling'' of a scattering volume of transverse
size $h$, leads to a loss of correlation of the dynamic structure factor
that can be characterized by a frequency 
$\nu_\text{tr} \propto v_0/h$.

The \emph{Deborah-number} is a measure for 
the importance of this effect compared to the decorrelation due to 
thermal diffusion taking place on a timescale $\left( D q^2 \right)^{-1}$. 
In this context, the Deborah number can be defined as,
\begin{equation} 
De = v_0\left(hDq^2\right)^{-1} 
\label{eq:Deborah}.
\end{equation}
Typical values for two values of $q$ are given in table~\ref{tab:samples}.

It was demonstrated \cite{Berne_Pecora,Edwards_JAP42,Pusey_JPD76,Chowdhury_AO84} that the 
contribution of the transit effect to the correlation function is reflecting 
the beam profile, being ``scanned'' by the flowing particles.
As the beam is defined by rectangular slits, the incident intensity  
follows a $\left( \frac{\sin(r)}{r} \right)^2$ distribution, 
the squared Fourier transform of a rectangle-function. 
Due to the guard slits, which are aligned to suppress the higher order peaks, 
the beam profile at the sample position can be well approximated by a 
Gaussian form, leading to 
\begin{equation}
  |g_{1,T}(t)|^2 \propto \exp \left[ - (\nu_\text{tr} t)^2 \right] .
  \label{eq:g1t}
\end{equation}

\subsection{Shear-induced effects}

The shear-induced contribution to the correlation function, 
$|g_{1,S}(\vq, t)|^2$ can be written as a sum over all pairs of particles 
or as a double integral over the scattering volume which can be 
approximated by a line of length $2R$, 
\begin{equation}
  |g_{1,S}(\vq, t)|^2 = \frac{1}{4R^2} \int _{-R} ^R \int _{-R} ^R  \cos \big( \vq t \vec{\delta v}(r_1, r_2) \big) \mathrm{d}r_1 \mathrm{d}r_2 .
\label{eq:dv}
\end{equation}
The integral can be performed analytically for a uniform shear rate 
(Couette geometry) \cite{Narayan_AO97}, leading to
\begin{equation}
|g_{1,S}(\vq, t)|^2 = \frac{\sin(q_\parallel v_0 t)}{q_\parallel v_0 t}.
\label{eq:sinx_x}
\end{equation}

In Appendix~\ref{appendix:shear}, Eq.~\eqref{eq:dv} was also solved analytically
for a different flow geometry, a Poiseuille flow described by a parabolic 
velocity profile Eq.~\eqref{eq:vf}, which provides a better description for 
the data presented 
here. The resulting shear-induced response to the correlation functions can be 
written using the error function of a complex argument,
\begin{equation}
  |g_{1,S}(\vq,t)|^2 = \frac{\pi^2 }{16 q_\parallel t v_0} \left| \erf \left( \sqrt{ \frac{4i q_\parallel t v_0}{\pi}}\right) \right|^2.
\label{eq:g1s}
\end{equation}

From Eq.~\eqref{eq:sinx_x} and \eqref{eq:g1s}, a shear-induced
frequency can be defined as
\begin{equation}
\Gamma _S={\bf v_0}\cdot {\bf q}
\label{eq:GammaS}.
\end{equation}
If the scattering alignment is supposed to be perfectly transversal, 
{\bf q}$\perp${\bf v}, $\Gamma _S$=0 and the XPCS correlation functions are
unaffected by the shear-induced effects. 

The relative importance of the shear-induced effects compared to thermal
diffusion is usually described by the \emph{Peclet number}
$P = \frac{\dot{\gamma} R^2}{D}$. 
As it can be seen in table~\ref{tab:samples}, the Peclet numbers are very high 
even at low shear rates, and the shear-induced effects are the 
dominating contribution to the measured correlation functions
if a non-transverse scattering geometry is used. 

Following the discussion above, it turns out that the relative influence of 
the shear can also be expressed in terms of a different dimensionless number
which will be called here \emph{Shear number}, given by the ratio between the
characteristic diffusion time $\tau _D=1/Dq^2$ and the shear-time,
$\tau _S=1/\Gamma _S=1/{\bf q}\cdot{\bf v_0}$,
\begin{equation}
S=\frac{ {\bf q}\cdot {\bf v_0}}{Dq^2}.
\label{eq:Snumber}
\end{equation}
Values for the shear number $S$ are given in table~\ref{tab:samples}
for a value of the longitudinal component of the scattering wavevector
$q_\parallel$=1.4$\cdot$10$^{-3}$~\AA$^{-1}$. The conclusion is the same as
the one resulting from the evaluation of the Peclet numbers -- the shear induced
effects are the dominating ones if a non-transverse scattering geometry is
used and it is hard or impossible to measure the thermal diffusion of the
particles in such a scattering geometry. 

\begin{figure}
\resizebox{0.75\columnwidth}{!}{%
  \includegraphics{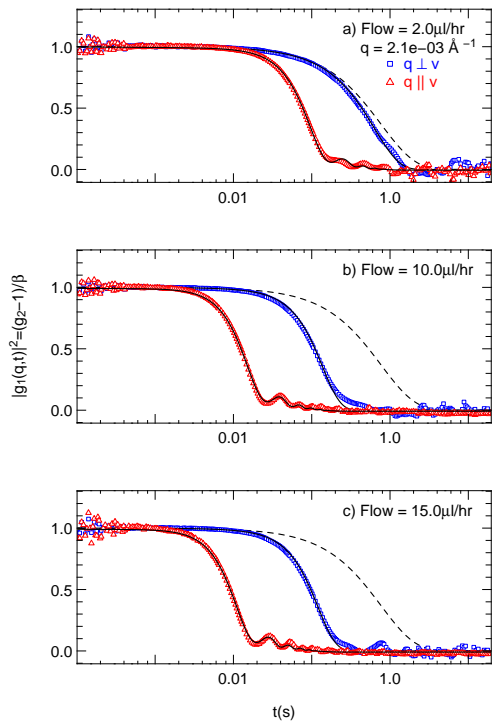}
}
\caption{Normalized correlation functions for a 
single wavevector, $q = 2.1 \cdot 10^{-3}\,\text{\AA}^{-1}$ and three different 
flow rates: a) $Q = 2$, b) $Q = 10$, and c) $Q = 15\,\mu\text{l/h}$. The 
dashed line shows the fit of a simple exponential to the zero-flow (reference) 
correlation function for the same value of $q$. The solid lines show fits 
with \eqref{eq:g2perp} and \eqref{eq:g2para} as described in the text.}
\label{fig:fits}       
\end{figure}

With the three components of the correlation functions determined by 
thermal diffusion, shear, and transit-time described by Eqs.
\eqref{eq:g1D}, \eqref{eq:g1s}, and respectively \eqref{eq:g1t}, the
normalized intensity fluctuation autocorrelation functions measured in a 
XPCS experiment with a sample undergoing shear flow can be written as,
\begin{multline}
\left|g_1({\bf q}, t) \right| ^2= \left( g_2({\bf q},t)-1\right)/\beta = \\
\exp \left[ -2 Dq^2 t \left( 1 -\frac{q_\perp q_\parallel}{q^2}\dot \gamma t 
  + \frac{q_\parallel^2}{q^2} \frac{(\dot\gamma t)^2 }{3}\right)\right] \cdot 
	\\ \exp \left[ - (\nu_\text{tr} t)^2 \right] \cdot
  	\frac{\pi^2}{16 q_\parallel t v_0} 
	\left| \erf \left( \sqrt{\frac{4 i q_\parallel t v_0}{\pi}} \right) \right|^2
  \label{eq:g2all}.
\end{multline}
As it will be shown in the following section, Eq.~\eqref{eq:g2all} fits 
very well
the measured correlation functions (Fig.~\ref{fig:fits})

\section{Results \& discussion}
\label{sec:Results}

In the $q$--range probed here, the static structure factor 
(Fig.\ref{fig:saxs}, inset) of the colloidal suspension 
in glycerol exposed to a shear flow is isotropic and independent of 
the flow rate. 
This means that the flow does not change the spacial probability 
distribution of the particles. However, the dynamic structure factor
(intermediate scattering function) is both flow-dependent and anisotropic.

Fitting the data obtained from a non-flowing sample with a simple 
exponential decay -- Eq.~\eqref{eq:g1diff}, as shown in 
Fig.~\ref{fig:noflow}~(inset), yields a mean diffusion coefficient of 
$D = 2.39 \cdot 10^{-15}\,\frac{\text{m}^2}{\text{s}}$. 
Using the Stokes-Einstein relation
\begin{equation}
  D = \frac{k_B T}{6 \pi \eta a} ,
  \label{eq:stokes-einstein}
\end{equation}
where $k_B$ is the Boltzmann constant and $T = 293\,\text{K}$ the temperature, 
the dynamic viscosity $\eta$ can be calculated to be 0.839\,Pa$\cdot$s which 
corresponds to a left-over water content of 3\% \cite{Handbook}.

%
%
%
%
%
%
%
%
%
%

\begin{table*}
  \caption{Characterizing quantities of the flow of the sample. As a function 
of flow rate $Q$, the mean velocity $v_0$, Reynolds-number $Re$, 
Deborah-number $De$ (for two $q$-values), shear rate $\dot{\gamma}$, 
Peclet number $P$, and Shear number \eqref{eq:Snumber} are shown.}
\label{tab:samples}       
\centering
  \begin{tabular}{l|rrrrrrrrr}
    \hline\noalign{\smallskip}
    $Q$ [$\mu\text{l/h}$] & 0.0 & 2.0 & 5.0 & 7.5 & 10  & 15  & 20  & 40 & 80 \\
    \noalign{\smallskip}\hline\noalign{\smallskip}
    $v_0$ [\microns/s] & 0.0 & 0.5 & 1.3 & 2.0 & 2.7 & 4.1 & 5.5 & 11 & 22 \\
    $Re$ [$10^{-6}$] & 0.0 & 1.7 & 4.2 & 6.3 & 8.4 & 13 & 17 & 33 & 67 \\
    $De$ ($q = 1.4 \cdot 10^{-3}\,\text{\AA}^{-1}$) & 0.0 & 0.12 & 0.30 & 0.44 & 0.59 & 0.89& 1.2 & 2.4 & 4.7 \\
    $De$ ($q = 5.3 \cdot 10^{-3}\,\text{\AA}^{-1}$) & 0.0 & 0.0083 & 0.021 & 0.031 & 0.041 & 0.062 & 0.083 & 0.17 & 0.33 \\
    $\dot{\gamma} \left[ \frac{10^{-3}}{s} \right]$ & 0.0 & 1.1 & 2.8 & 4.2 & 5.6 & 8.3 & 11 & 22 & 44 \\
    $P$ [$10^6$] & 0.0 & 0.46 & 1.2 & 1.7 & 2.3 & 3.5 & 4.6 & 9.3 & 19 \\
    $S$ ($q_\parallel = 1.4 \cdot 10^{-3}\,\text{\AA}^{-1}$) & 0.0 & 36 & 93 & 143 & 193 & 293 & 393 & 786 & 1429 \\
    \noalign{\smallskip}\hline
  \end{tabular}
\end{table*}

Using the obtained diffusion coefficient and viscosity, the flow can be 
completely characterized (see relevant values in table~\ref{tab:samples}). 
It can be concluded that the Reynolds-numbers are several orders of 
magnitude below the onset of turbulent flow. The 
Deborah-number -- Eq.~\eqref{eq:Deborah} is  
$q$-dependent. At the smallest $q$--values measured, it is not possible any 
more to assume $De \ll 1$ (even at low flow rates) and the transit-time
effects must be considered for an accurate description of the correlation
functions. At higher values of $q$, the Deborah numbers $De$ become much
smaller and the transit-time effects are less important.
Increasing the beam size $h$ offers another practical possibility to lower 
the Deborah number, if working with a lower speckle
contrast is acceptable. 
The Peclet number $P$ and/or the q-dependent shear numbers S are much 
larger than unity even at the smallest flow rates,
showing that the shear-induced effects are always important.

The intensity fluctuation correlation functions were measured for several
values of the scattering wavevector {\bf q} in both transverse
(\qperpv) and longitudinal (\qparav)
scattering geometries and for several flow rates between 0 and 80~$\mu$l/h
and fitted using Eq.~\eqref{eq:g2all}.  
For a transverse scattering geometry $q_\parallel=0$, and Eq.~\eqref{eq:g2all} 
leads to
\begin{equation}
  g_{2,\perp}(q,t) = 1 + \beta \cdot \exp \left[ -2 \frac{t}{\tau} \right] \cdot \exp \left[ - (\nu_\text{tr} t)^2 \right]
  \label{eq:g2perp},
\end{equation}
with the diffusion time $\tau=1/Dq^2$ and the transit-induced frequency 
$\nu_{tr}$ as fitting parameters (in addition to the speckle contrast $\beta$
and the baseline value). The diffusion coefficient was obtained by fitting
the low-flow transverse, \qperpv\ correlation functions and was
subsequently fixed in the high-flow data which were used to determine 
the flow-dependent, transit-induced frequency $\nu_\text{tr}$. 
Both parameters $\tau$ and $\nu_\text{tr}$ were fixed to the values obtained from
transverse scans when fitting the \qparav\ data with
\begin{multline}
  g_{2,\parallel}(q,t) = 1 + \beta \cdot \exp \left[ -2 \frac{t}{\tau} 
\left( 1 + \frac{(\dot{\gamma}t)^2}{3} \right) \right] \\
  \cdot \exp \left[ - (\nu_\text{tr} t)^2 \right] 
  \cdot \frac{\pi \left| \erf \left( \sqrt{i \Gamma_S t } \right) \right|^2}
	{4 \Gamma_S t}
  \label{eq:g2para},
\end{multline}
from which a fitted shear-induced frequency $\Gamma_S$ was obtained.

\begin{figure}
\resizebox{0.55\columnwidth}{!}{%
  \includegraphics{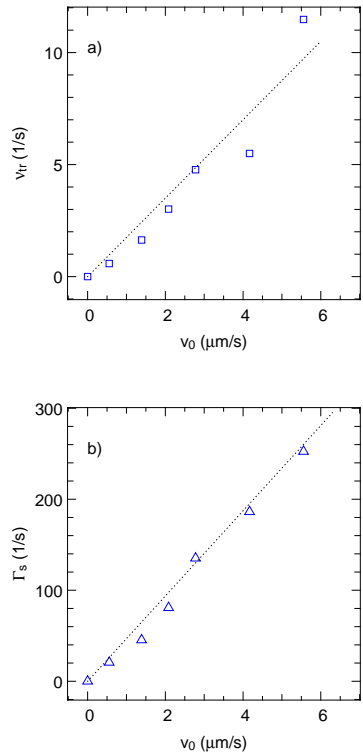}
}
\caption{The fitted parameters versus the flow velocity of the sample; 
a) shows the results of fitting the transverse correlation functions with
Eq.~\eqref{eq:g2perp}. Fits with Eq.~\eqref{eq:g2para} of the longitudinal 
correlation functions yields the values for the shear-induced frequency $\Gamma _S$ displayed in b).}
\label{fig:fitparams}       
\end{figure}

Example fits for a single value of $q$ and three different flow rates 
are shown in Fig.~\ref{fig:fits}. The parameters resulting from the 
fitting procedure described above, $\nu_\text{tr}$ 
and $\Gamma _S$, are plotted in Fig.~\ref{fig:fitparams} as a function of the 
nominal flow velocity $v_0$ computed from the volume flow rate. The 
shear rate $\dot\gamma$ was left in place in 
Eq.~\eqref{eq:g2para} only for reasons
of logical consistency with Eq.\eqref{eq:g2all} and \eqref{eq:g1D}, but its 
influence on the shape of the correlation functions is too small (for any
reasonable values of the shear rate) to allow any meaningful fitting. 
As it can be seen from Fig.~\ref{fig:fits}, the quality of the fits 
is quite remarkable, especially for \qparav\ and
for the higher flow rates, even if the only dominant parameter for these
fits is $\Gamma_S$, the shear-induced frequency. 

From a practical point of view, in order to be able to measure the 
diffusive component of the particle dynamics under flow, its 
relative importance must
be high compared with that of the shear and transit-time induced effects.
Keeping a transverse scattering geometry is the only method to limit the
shear-induced effects, but in practice any small misalignment becomes the
limiting factor at high-enough flow rates. If $\phi$ is the angle 
between {\bf q} and the direction orthogonal to {$\bf v_0$}, the shear-induced
relaxation rate -- Eq.\eqref{eq:GammaS} -- becomes \cite{Salmon},
\begin{equation}
\Gamma_S=v_0 \phi q
\label{eq:GammaSphi}.
\end{equation} 

\begin{figure}
\resizebox{0.6\columnwidth}{!}{%
  \includegraphics{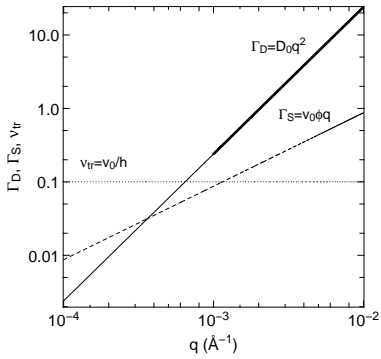}
}
\caption{Dispersion relationships for the diffusion ($\Gamma_D=D q^2$), 
shear ($\Gamma _S= v_0 \phi q$), and transit ($\nu_{tr}=v_0/h$) relaxation 
rates. The (example) values used here to estimate $\Gamma_D$, $\Gamma _S$, 
and $\nu _{tr}$ were $D_0\approx 2.4\cdot10^5$~\AA/s, $\phi\approx$
0.01 (0.5 deg), $v_0=1 \mu$m/s. The thick solid line shows the q region 
where the thermal diffusion dominates the dynamic signal and can be measured
by XPCS.}
\label{fig:DifShTr}       
\end{figure}

To illustrate the impact of even small misalignments,  
the dispersion relationships for thermal diffusion $\Gamma_D=Dq^2$, 
shear-induced effects, Eq.\eqref{eq:GammaSphi}, with $\phi$ assumed to be
0.01 (corresponding to a misalignment of 0.5~deg) and $v_0$=1~$\mu$m/s,  
are plotted in Fig.~\ref{fig:DifShTr} together with the 
(q-independent) dispersion for the transit-time effects. In order to be able
to measure the thermal diffusion of the particles, $\Gamma_D$ must dominate 
over the other, flow-induced relaxation rates. For the values chosen in this
example, this condition is fulfilled in the q-range for which the diffusion 
dispersion relationship is highlighted. 
From Fig.~\ref{fig:DifShTr} it is clear that for any flow velocity larger
than a few $\mu$m/s, the shear-induced relaxation rates $\Gamma_S$ will shift
upwards which would ``push'' the possible q-range to higher values. 
Equivalently, working at a fixed value of $q$ in order to obtain specific
information about the dynamic properties on a certain length-scale 
puts a condition on the maximum flow rate for which the correlation 
functions are still dominated by their diffusive components. 
At the same time, it is also clear that in the case of
light-PCS, with smaller values of $q$, these conditions are harder to fulfill
\cite{Ackerson_Clark_JPhysique81}.  

From this analysis, it results that fitting the correlation functions
obtained in transverse scans with Eq.~\eqref{eq:g2perp} is justified only 
at the lowest flow rates probed here (assuming that it is difficult to achieve
a vertical alignment better than $\phi$=0.01). For such low flow rates, the 
correlation functions will not depend much on neither the finite transit-time
nor on the shear time and fits with a simple exponential form will
render the correct diffusion time and/or diffusion coefficient.

\begin{figure}
\resizebox{0.6\columnwidth}{!}{%
  \includegraphics{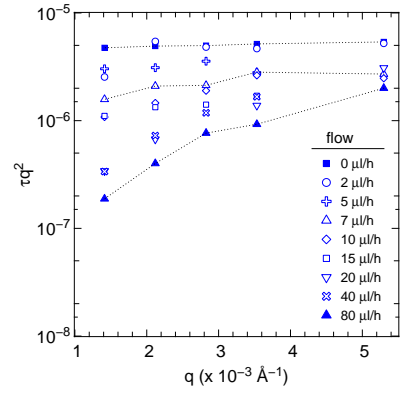}
}
\caption{Correlation times (multiplied by $q^2$) as a function of $q$.
At low flow rates the correlation times are unaffected by the transit time 
and/or the shear and the diffusive dynamics of the particles can be measured
in a flowing solution.}
\label{fig:tauq2}       
\end{figure}

The correlation times in Fig.~\ref{fig:tauq2} were obtained by performing 
fits to the transverse correlation functions with simple exponential forms.
While at most of the higher flow rates probed here the influence of the 
flow on the correlation times is clear, at the low-enough flow rates 
(Q$\leq$2$\mu$l/h), the correlation times are still flow-independent 
and follow the $\Gamma\propto q^2$ dispersion relationships expected for 
Brownian motion. 

Therefore, using XPCS on a flowing sample seems applicable to measure 
the ``intrinsic'' dynamics of the sample. While the scattering functions 
recorded with \qparav\ are completely dominated by shear effects, these do 
not contribute (or contribute much less) to the spectra at \qperpv.

\section{Summary}

\label{sec:Summary}

The correlation functions from a colloidal suspension undergoing shear flow
were measured by homodyne XPCS. The data could be explained by computing 
the intermediate scattering function as a product of the scattering 
functions due to diffusion, transit, and shear.

Our measurements suggest that it should be possible to measure samples 
under flow and obtain the diffusion coefficient and/or particle size 
information, etc. For shear and Deborah-numbers $< 0.1$, the correction 
is negligible and equation~\eqref{eq:g1diff} can be used. For higher Deborah
and/or shear numbers, a more complicated convolution between different 
dynamic mechanisms takes place and the correlation functions are dependent
on at least three important factors -- thermal diffusion, transit-time, and 
shear. 
Measurements at shear and/or Deborah numbers above $\approx 1$ are completely 
dominated by shear-induced or (in some particular cases) transit time effects 
and not suitable any more for a precise extraction of diffusive properties. 
However, as both the $De$ and $S$ numbers are $q$-dependent, it is possible 
to tune the settings within some limits to access also higher flow rates.

In practice, for a specific alignment a ``calibration'' procedure (i.e.
probing the dynamics of a colloidal suspension as it is done here), should
provide a graph such as the one shown in Fig.~\ref{fig:DifShTr}, which
provides a lower limit for $q$ accessible for a given flow rate or,
equivalently, the maximum flow rate that can be set while still being able
to measure the diffusive dynamics at a specific value of $q$ with the 
desired degree of accuracy.

This method can therefore be used to measure the dynamical properties of 
many soft and biological samples which suffer from beam damage, or to
perform time-resolved studies in mixing devices. 

\section{Acknowledgments}

We acknowledge many useful discussions with Anders Madsen,
Abdellatif Moussa\"id, Federico Zontone, Chiara Caronna, Jean-Baptiste Salmon, 
Fanny Destremaut, and help in designing the experiment and building the
flowcells from Henry Gleyzolle. The work of SB and THJ was part of the student
research internship program at the ESRF. They wish to acknowledge 
financial support from the ESRF and administrative support and guidance 
from Catherine Stuck and the HR department and from the ID10A staff.

\appendix \section{Shear-induced correlation}
\label{appendix:shear}

In the following, Eq.~\eqref{eq:dv} is solved for a Poiseuille flow,
with a velocity profile described by~\eqref{eq:vf}, and a 
closed-form expression for the shear-induced
correlation $|g_{1,S}(\vq, t)|^2$ is calculated. 

The velocity difference between two particles located 
at different positions across the flow channel $r_1$ and $r_2$ can be 
written as $\vec{\delta v}(r_1, r_2) = 2 \vec{v_0} 
\frac{r_1^2 -r_2^2}{R^2}$, the integrals over the two independent variables 
$r_1$ and $r_2$ are equivalent and, after applying trigonometric addition 
identities, the double integral in equation~\eqref{eq:dv} can be reduced to the 
integration over a single variable r,
\begin{multline}
  |g_{1,S}(\vq, t)|^2 =
  \frac{1}{4R^2} 
  \left[\int _{-R}^R \mathrm{d}r \cos \left( 2 q_\parallel t v_0 \frac{r^2}{R^2} \right) \right]^2+\\
  \frac{1}{4R^2} \left[ \int _{-R}^R \mathrm{d}r \sin \left(2 q_\parallel t v_0 \frac{r^2}{R^2} \right) \right]^2.
\end{multline}
These integrals can be reduced to Fresnel integrals \cite{NRC},
\begin{equation}
  |g_{1,S}(\vq, t)|^2 = \frac{\pi}{4 q_\parallel t v_0}
  \left[ C^2 \left( \sqrt{\frac{4 q_\parallel t v_0}{\pi}}\right) + 
  S^2 \left( \sqrt{\frac{4 q_\parallel t v_0}{\pi}}\right) \right] ,
\end{equation}
which are related by (using the imaginary number $i$=$\sqrt{-1}$~)
\begin{equation}
  C(v)+iS(v) = \frac{1+i}{2} \erf \left( \frac{\sqrt{\pi}}{2}(1-i)v \right),
\end{equation}
to the error-function,
\begin{equation}
  \erf(u) = \frac{2}{\sqrt{\pi}} \int_0^u e^{-t^2} \mathrm{d}t .
\end{equation}
A normalized expression for the shear-induced correlation function 
$|g_{1,S}(q,t)|^2$ is, therefore given by
\begin{equation}
  |g_{1,S}(\vq,t)|^2 = \frac{\pi^2 }{16 q_\parallel t v_0} \left| \erf \left( \sqrt{ \frac{4i q_\parallel t v_0}{\pi}}\right) \right|^2.
\end{equation}

\bibliographystyle{epj.bst}
\bibliography{references}

\end{document}